# Event Correlation and Forecasting over Multivariate Streaming Sensor Data

Vassilis Papataxiarhis[1*], Stathes Hadjiefthymiades[1]

[1] *Pervasive Computing Research Group, Department of Informatics and Telecommunications, National and Kapodistrian University of Athens, GR – 157 84, Athens, Greece*

**Abstract:** Event management in sensor networks is a multidisciplinary field involving several steps across the processing chain. In this paper, we discuss the major steps that should be performed in real- or near real-time event handling including event detection, correlation, prediction and filtering. First, we discuss existing univariate and multivariate change detection schemes for the online event detection over sensor data. Next, we propose an online event correlation scheme that intends to unveil the internal dynamics that govern the operation of a system and are responsible for the generation of various types of events. We show that representation of event dependencies can be accommodated within a probabilistic temporal knowledge representation framework that allows the formulation of rules. We also address the important issue of identifying outdated dependencies among events by setting up a time-dependent framework for filtering the extracted rules over time. The proposed theory is applied on the maritime domain and is validated through extensive experimentation with real sensor streams originating from large-scale sensor networks deployed in ships.

## 1. Introduction

In a Sensor Network (SN), sensor nodes constitute the network components that collect data about the environment within which they are deployed. Sensing elements capture contextual information to support context-aware applications from diverse domains (e.g., maritime, intelligent transportation, smart farming). Being based on the collected information, a SN intends to detect events that take place during the network lifetime and may depict, or even affect, the state of the system. The term 'event' is used to describe an alteration on one or more variables monitored by the system (we call them context attributes - CA).

In complex SNs, two main kind of processing modes can be distinguished with respect to events (Fig. 1). *Online event processing* focuses on real- or near real-time event detection, identification of time-dependent correlations and causalities, prediction of upcoming system states and filtering of outdated data. On the other hand, *offline processing* includes, among others, event storage, post-processing of stored events and data-warehousing, and visualization facilities that support the decision makers to proceed with remedial actions if needed (e.g., modify system parameters, substitute a sensor which is out of order). This paper focuses on the online processing of events originated from sensor networks. We follow a stepwise approach to analyze the CA and unveil hidden interconnections among the several event types occurred during the operation of a sensor network.

Sensors streams mostly arrive as streams of raw data that provide instant measurements or summaries (e.g., mean, max) regarding observed phenomena that may change over time. Due to its frequency, raw data are of limited value even for system experts that intend to reach a higher level of understanding for the internal dynamics of the system. The sensor streams are handled and analyzed in an online fashion in order to identify times when the probability distribution of the respected time series changes. Each sensor stream is transformed into a binary stream that represents time points of unexpected behaviour of a CA (**event detection**). Event streams can provide refined information of the data that may be of high importance to the decision makers that need to act proactively in order to sustain the proper behaviour of the system. For example, the abnormal increase of the values reported by a smoke detector could essentially be an evidence of a fire. As a next step, we focus on the derivation of a dependency structure that describe time-dependent regulations between the incoming

[*] Corresponding author.
  Email addresses: vpap@di.uoa.gr (V. Papataxiarhis), shadj@di.uoa.gr (S. Hadjiefthymiades)

data streams and illustrate possible transitions on the system state (**event correlation**). The resulted rules are extracted from dependency structures that are constructed in near real-time to represent interdependencies among the measured variables **(event prediction)**. The derived dependencies that result in separate time steps are filtered in order to balance between outdated patterns and event sequences of previous steps that could still be of importance **(event filtering)**. Fig. 2 presents the workflow for the online processing of events originated from sensor networks as we consider it in the context of this work.

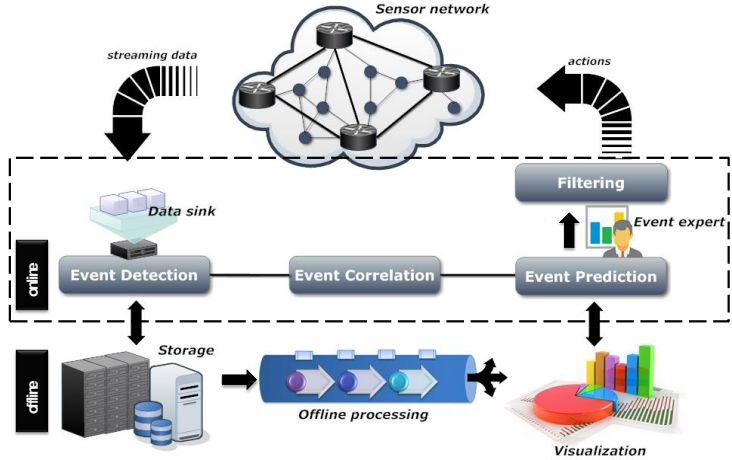

**Fig. 1.** Event management lifecycle in sensor networks.

The contribution of this study is multifold. First, we show that the determination of events in sensor networks can be accommodated within a change detection framework in the context of multivariate time series. Then, we propose an event correlation scheme based on the idea of partial matching [1] by essentially implementing a variable-order Markov model for the correlation of multivariate event data. Next, we discuss a framework for modeling event dependencies through a probabilistic temporal logic programming paradigm. Finally, we propose a methodology to deal with outdated dependencies.

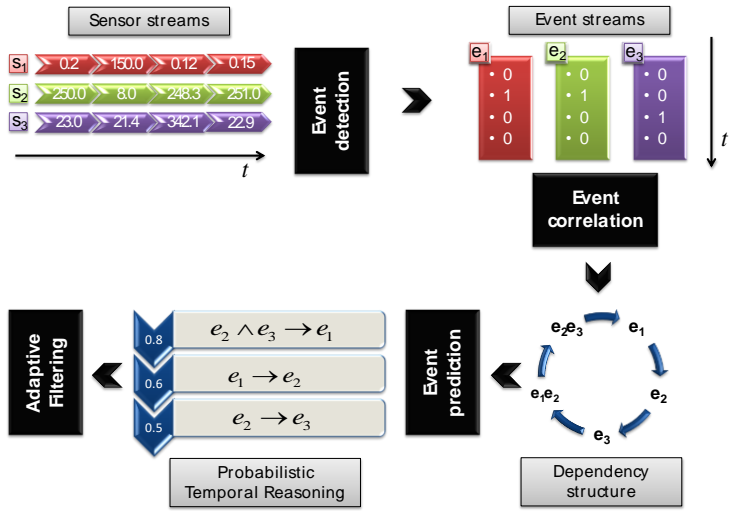

**Fig. 2.** Online processing of events in sensor networks.

The rest of the paper is organized as follows; Section 2 provides a review of the research efforts related to the areas of online event correlation and forecasting; Section 3 discusses change detection schemes; Section 4 proposes a novel event correlation method for multivariate event data; Section 5 illustrates the use of probabilistic temporal logics to reason over event data; Section 6 describes a framework for adaptive filtering of event dependencies; The discussed approaches are evaluated in

Section 7; Section 8 concludes the paper with open issues and next research steps.

## 2. Prior work

Event management has gained much popularity among the research as well as the industrial community for many decades. From a research perspective, literature provides a wide range of approaches for the detection and correlation of events in real-time data management systems. In [2], the authors propose a real-time probabilistic framework for the detection of abnormalities in streaming event logs that come from multiple event sources. The proposed scheme assumes the events already detected and the identification of abnormalities is based on the derivation of a directed probabilistic graph that represents the historical footprint of the event occurrences. The graph vertices (i.e., event types) are specified by the prior probabilities for the occurrence of event types while each directed edge comes with a conditional probability between the connected event types. The authors also propose a metric to quantify the similarity between two event streams. A change is detected whenever this metric exceeds a threshold comparing the graph resulted from the recent event records (i.e., stream tail) to the complete graph of the original stream. A similar graph structure is presented in [3] with the edges corresponding to the joint probabilities of two event types (i.e., bidirectional graph). The authors apply measures originated in information theory in order to provide a historical analysis of abnormality events (i.e., events are considered known and detected from historical logs). They also provide metrics and a stepwise analysis for the determination of root-cause events as well as events that can, possibly, lead to a system crash. Both frameworks do not take into account sequences of correlated events since the derived conditional and joint probabilities, respectively, refer to events occurred during the same time window, but not in subsequent time steps. For example, consider a case of an event stream where whenever an event of type A occurs at time $t$, an event of type B also occurs at time step $t+1$. Similarly, implications imposed by joint occurrences of events are not considered due to time complexity (e.g., event types A and B have to occur both in order to result in the occurrence of event type C). In [4], the author focuses on change detection criteria in multidimensional streaming data. The paper investigates well-known metrics for change detection, such as Kullback-Leibler distance [5] and Hotelling's T-square test for equal means [6], providing a log-likelihood justification. It also proposes and evaluates a semi-parametric log-likelihood criterion that perform better in multidimensional data. In [7], the authors propose a probabilistic scheme for the correlation of distributed events in the network security field. Specifically, they apply a HMM [8] and Kalman filtering [9] to unveil spatial and temporal correlations among the observations and the hidden states of internet security attacks. Kalman fitering and Cumulative Sum algorithm [10] are adopted to detect changes in real time in [11]. The paper also evaluates several criteria for change detection in streaming data.

Industrial maintenance has also received significant attention in academic literature for many decades. Specifically, a lot of studies have been devoted in the area of *condition-based predictive maintenance* [12] for the prediction of faults and errors in complex technical systems including sensor networks. Most of industrial approaches are based on existing statistical methods to identify alerting situations due to abrupt changes of the probability distributions of the observed parameters [13]. In [14], the authors discuss the problem of rare events prediction in multidimensional data and they propose two approaches that address the problem. The first approach focuses on degradation detection (i.e., abnormal system behavior) that is based on one-class Support Vector Machines model [15]. The proposed method formulates a quadratic program that detects anomalies by minimizing the distance of each new multidimensional data vector with regard to a training dataset. Then, a moving average model is used for modeling multidimensional degradation behavior in time. The second approach exploits a regularized logistic regression classification method where the predictor function consists of a transformed linear combination of explanatory variables. Both approaches are evaluated over datasets resulted by real-world cases from aircraft operational performance area. In [16], the authors propose one model-based and one data driven method for predicting faults in multi-sensor systems. The former approach is based on an autoregressive moving average model where the estimated model parameters are further used for implementing the change detector that is realized as a Neyman-Pearson hypothesis test [17]. The second approach takes advantage of a HMM and Viterbi algorithm [18] to estimate the most likely sequence of hidden states. Both approaches are validated through simulations with synthetic data.

## 3. Event detection

In this section, we discuss the process of event detection and the generation of event streams over an existing set of sensor streams. The problem of detecting events over multiple data streams can formally be expressed as follows. We observe in real-time, with some uniform frequency, multivariate time series of quantitative system performance parameters. Let $s_i$ be a *data stream* reporting numeric values and let $s_i(t)$ be the value of stream $s_i$ at time $t$, $t \geq 0$. Assuming that $n$ data streams are synchronized to report their values periodically, we represent the set of multidimensional contextual information at each time $t$ by the *context vector (CV)* $CV_t = (s_1(t), s_2(t), ..., s_n(t)) \in \Re^n$. Practically, each stream formulates a univariate time series while the context vector stream represents a multivariate time series. Event detection over sensor data aims to determine the values $s_i(t)$ that constitute abrupt changes within a context vector stream. Specifically, each context vector of length $n$ is transformed into a binary vector of the same length with each value representing a possible change in the respective data stream. Such deviations of the normal behaviour are considered as *events* and the binary vector as *event vector (EV)*. In particular, an event can be an observation, which does not conform to an expected pattern in the data set. The events may be caused by several reasons including sensor faults or malfunction, outliers or substantial changes that may affect the behaviour of the system. Hence, an event vector at time $t$ is represented by $EV_t = (e_1^t, e_2^t, ..., e_n^t) \in \{0,1\}^n$, where $e_i^t = e_i(t)$ is the binary value representing whether an abnormal behaviour took place for stream $s_i$ (value '1') at time $t$ or value $s_i(t)$ was included in the expected range of values.

The transformation of a CV into an EV is based on change detection algorithms that aim to identify abnormal deviations of the current values with respect to the values arrived in previous steps. Change detection algorithms can be distinguished into the following categories:

- *Univariate change detection*. The algorithms belonging in this category (e.g., CUSUM [10], Shewhart control charts [19]) take into consideration each data stream separately and detect possible abnormalities through a sequential time series analysis. These approaches assume that changes within a certain stream do not affect other streams.
- *Multivariate change detection*. The algorithms of this category (e.g., ARfit [20] [21]) exploit multivariate autoregressive (MAR) models [22] to represent each context vector as a linear sum of the previous activity. Then, obtaining a binary value that indicates the change (or non-change) for a specific variable (i.e., data stream) is reduced into a single thresholding operation between the estimated future vector and the actual vector arrived.

Two univariate change detection algorithms are examined in the context of this paper.

*3.1 Cumulative Sum (CUSUM) Algorithm*

The cumulative sum (CUSUM) algorithm [10] attempts to detect a change on the distribution of a time series with respect to a target value at real-time. Specifically, we consider a univariate time series $x_t \in \mathbb{R}$ consisting of data values collected over time and a target value $\mu$ for this data stream. CUSUM involves the calculation of positive and negative changes ($P$ and $N$, respectively) in the time series $x_t$ cumulatively over time and it compares these changes to a positive and a negative threshold ($thresh^+$ and $thresh^-$, respectively). Whenever these thresholds are exceeded, a change is reported through the above-detection and below-detection signals ($s^+$ and $s^-$, respectively) while the cumulative sums are set to zero. To avoid the detection of non-abrupt changes or slow drifts, the algorithm takes into consideration tolerance parameters for positive and negative changes ($k^+$ and $k^-$, respectively).

The algorithm assumes that the time series follow a normal distribution. In order the algorithm to work properly, the tolerance and threshold parameters should be tuned in a way that determines what an actual change is for a specific time-series [23].

Fig. 3 illustrates an example of the CUSUM algorithm over a data stream $x_t \in \mathbb{R}$ where two changes (a positive and a negative one) are detected. The figure presents the original data stream and the time steps where a change is detected by the algorithm.

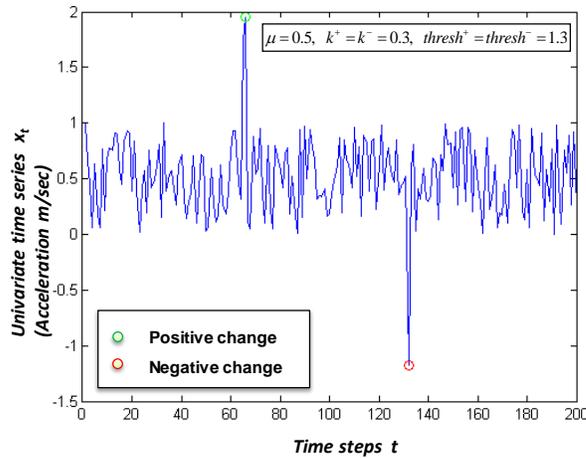

**Fig. 3.** Original data stream measuring acceleration by using sunspot sensors [24] and change detection based on CUSUM algorithm.

*3.2 Shewhart Controller*

Shewhart control charts [19] provide a statistical measure to detect abrupt shifts of a univariate time series. In the Shewhart control chart, a variable $x_t \in \mathbb{R}$ is detected to deviate at time $t$ from its normality whenever it exceeds one of the control limits specified by the algorithm: the Upper Control Limit (UCL) and the Lower Control Limit (LCL). The control limits are defined as the distance from the current mean value of the statistical process $x_t \in \mathbb{R}$.

Fig. 4 illustrates an example of the Shewhart controller over a sensor stream (i.e., Arduino accelerometer [25]) $x_t \in \mathbb{R}$. Similarly to CUSUM algorithm, Shewhart control chart has to be applied to each variable separately in case of multivariate time series. However, the algorithm does not assume normal distribution for $x_t$. This makes the algorithm quite robust in real-life datasets where most of the times there is no knowledge available for the probability distribution that a data stream follows. On the other hand, the algorithm is less adaptive compared to CUSUM since the control limits may be barely modified in case of long time series.

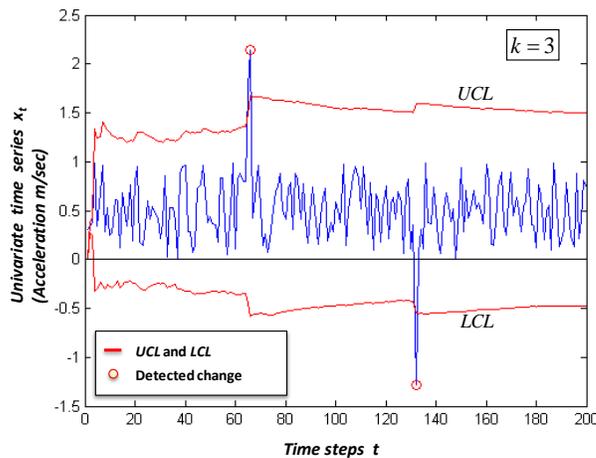

**Fig. 4.** Original sensor stream measuring acceleration through MPU 60-50 [25] and Shewhart change detection chart.

4. **Event correlation**

Over the past decades, a lot of effort has been devoted in the correlation of event data originated from IT infrastructures in order to extract patterns that depict the internal dynamics of systems and act proactively. Typical event correlation schemes that operate over univariate time series perform under the following assumptions:
- A transition from object (i.e., event or sequence of events) *A* to object *B* occurs if and only if *B* occurs immediately after *A* (i.e., not within a time window).
- Only one object is considered at each step of the sequence (i.e., there are no objects occurring at the same time).

However, things may differ in case of event correlation over multivariate sensor data. First, there is no guarantee that events would happen one at a time. On the contrary, an alerting situation or a malfunctioning system is expected to lead to several events triggered at the same time step. For example, consider the case of a fire that is monitored through multiple smoke detectors, temperature and humidity sensors. Most of the sensors that monitor an affected area are expected to experience a significant drift in the probability distribution of their reported values, thus leading to multiple events within the same time steps. Further to the above, since events occur rarely when no alerting situation takes place, the probability of having no events appearing at specific time steps should be also considered.

Here we try to address the above challenges in the context of multivariate streaming event data. We consider the occurrence of event vector streams as a stochastic process and we propose a variable-order correlation scheme based on the idea of *partial matching* [1], which had initially been applied for web prefetching purposes and we have adapted it to the context of multivariate time series.

*4.1 Variable-order correlation scheme*

In the context of web prefetching, the partial matching algorithm [1] faces the problem of predicting the upcoming $l$ URL accesses of a client based on information about the past $m$ steps. The algorithm maintains a data structure that keeps track of the past URL sequences of length up to $m+l$. To find candidate URLs for prefetching, the algorithm determines all subsequences in the history structure that match any suffix of the last $m$ accesses. A cut-off threshold $p_{thr}$ is used to discard URLs with low probability of occurrence. All URL items that match these suffixes and their probability exceeds the cut-off threshold $p_{thr}$ are prefetched.

Here we propose a variable-order event correlation algorithm by adapting the idea of partial matching to the context of multivariate streaming data. The algorithm essentially implements the variable-order Markov model idea [26] where Markov chains of orders $1,...,m$ are combined to predict event sequences of length up to $m$. Similarly to the original partial matching algorithm, this approach approach takes into account the following two parameters :
(a) Number of past multivariate event vectors considered ($m$). This parameter represents the *maximum order* of the model.
(b) Number of steps that the algorithm would predict in the future ($l$).

Let $I = \{e_1,...,e_n\}$ be the set of the considered event types $e_i$ with regard to the respective sensor streams $s_i$, $i \in [1,n]$. We represent the $n$–dimensional event vector at time $t$ by $EV_t = (e_1^t, e_2^t,..., e_n^t) \in \{0,1\}^n$, where $e_i^t = e_i(t)$ are the binary values resulted from the change detection step and they represent whether an abnormal behavior took place for stream $s_i$ (value '1') at time $t$ or the value $s_i^t = s_i(t)$ that arrived was in the expected range.

The algorithm operates in an online fashion by maintaining a data structure that keeps track of the past sequences of events of length up to $m+l$. The considered data structure is a collection of trees $T = \{T_{I_r}\}$, $I_r \in \mathbf{P}(I)$, instead of multiple graphs, in order to conserve space since common prefixes of multiple sequences are stored only once. The index $I_r \subseteq I$ of each tree represents the event subset related to the root node of the tree. At each time step $t$ where a new event vector $EV_t = (e_1^t, e_2^t,..., e_n^t) \in \{0,1\}^n$ arrives, the algorithm updates the data structure so that all event sequences of length up to $m+l$ are included. Hence, the height of each tree is $h_{max} = m+l-1$ at most. Each *tree node* $v$ can be represented as a tuple of the form:

$$v = \langle I_v, N_v^t \rangle, \ I_v \subseteq I \tag{1}$$

where $I_v$ is a specific set of events related to this node and $N_v^t$ is the frequency of the observed pattern starting from the root node until the current time step $t$. At each step, the frequencies referring to sequences of the last $m+l$ steps are updated. Contrary to the stepwise correlation algorithm, the nodes represent combinations of events that occurred at the same time step, not states of all possible events realizations. Hence, multiple nodes may correspond to the same time step. The reason behind this decision is the practical need of keeping different frequencies for the same sub-patterns that occur within bigger event patterns (e.g., sub-pattern BC may occur within patterns ABC and CBC).

An indicative example of the algorithm with $m=2$ and $l=1$ is presented in Fig. 5 (steps 1-3) and Fig. 6 (steps 4-5). The event vectors refer to three types of events (A, B and C). At each step, red rectangles illustrate root nodes for new trees that have to be constructed. Black rectangles represent nodes that remain unchanged with respect to the previous step. Purple rectangles depict nodes that existed in the previous step but they have to be updated. New leaf nodes are highlighted in cyan. The integer number close to each node denotes the frequency of the observed pattern starting from the root of the tree.

The prior probability of an event set $I_r$ of a root node equals to the frequency $N_r^t$ of the root node divided by the current time step and is represented by $P_r^t = \dfrac{N_r^t}{t}$. For example, the prior probability of the event $C$ in step 5 equals to $P_C^5 = \dfrac{2}{5}$. A path within a tree represents a sequence of events that has been occurred and it is annotated with a probability value. The probability of such an event pattern of the form $r,...,u,v$ (i.e., path within a tree starting from root node $r$) equals to the frequency of the node $v$ at the current time step divided by the frequency of the node $u$ one time step before and is represented by $P_{r,...,u,v}^t = \dfrac{N_v^t}{N_u^{t-1}}$. For example, the probability of pattern $A \rightarrow B \rightarrow \varnothing$ after time step 5 equals to $P_{A,B,\varnothing}^5 = \dfrac{N_\varnothing^5}{N_B^4} = 1$.

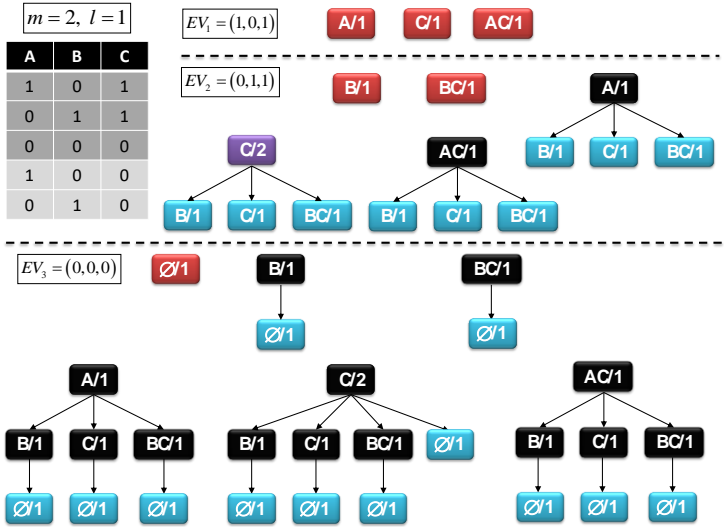

**Fig. 5.** Variable-order event correlation example – Steps 1-3.

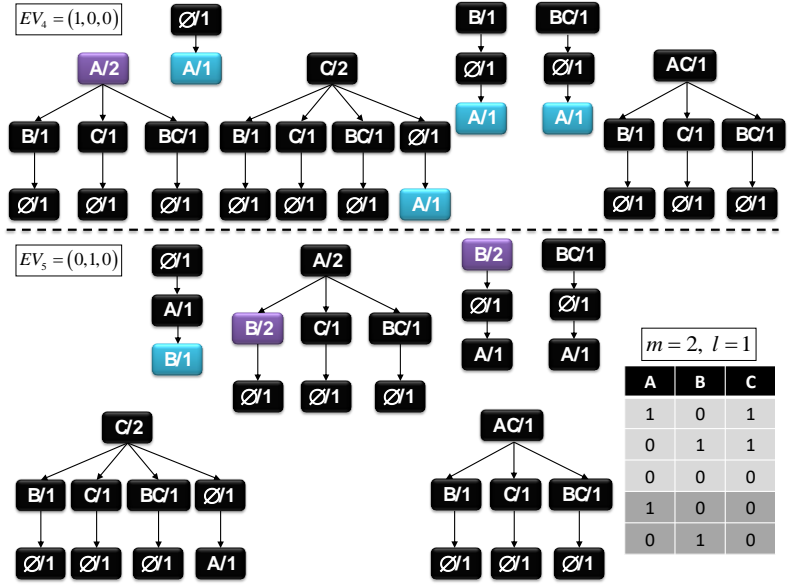

**Fig. 6.** Variable-order event correlation example – Steps 4-5.

The variable-order correlation algorithm assumes, in the worst case, that all possible combinations of $n$ different types of events must be kept at each step. In that case, the number of possible trees in the worst case is $\sum_{i=1}^{n}\binom{n}{i} = 2^n$. However, storing all possible states is of little use in practice since the probability of occurrence of a big number of events at the same time step is practically very low. If we consider only combinations consisting of $k$ events at most (with $k$ fixed and independent of $n$), then the upper limit on the number of the considered nodes is significantly reduced to the following estimation:

$$|V| = \sum_{i=0}^{k}\binom{n}{i} = \binom{n}{0}+\binom{n}{1}+\binom{n}{2}...+\binom{n}{k} = 1+n+\frac{n\cdot(n-1)}{2}+...+\frac{\prod_{j=0}^{k-1}(n-j)}{k!(n-k)!} = O(n^k) \quad (2)$$

This means that, assuming that combinations of $k$ event types at most (with $k$ fixed) will occur, the time complexity of the algorithm becomes polynomial. Since the event frequencies can be updated cumulatively, the algorithm is able to construct its history structure on the fly by continuously updating it for every new event vector arrived. Additionally, under the same assumption, the maximum number of nodes included in each tree (i.e., in case that all trees are full and complete) is

$\sum_{i=1}^{h_{max}}\left(\sum_{j=1}^{k}\binom{n}{j}\right)^i = \frac{1-n^{k(l+m-1)}}{1-n^k} = O(n^{m+l-1})$ where $h_{max} = l+m-1$ is the maximum height of each tree.

Practically, this means that the total number of nodes that need to be updated in each step for all the trees is polynomial ($O(n^{k+m+l-1})$), given that the algorithm will consider combinations of $k$ event types at most. Note that, similarly to the stepwise correlation approach, all frequencies can be updated cumulatively with the expansion of the stream.

## 5. Event prediction

The ability to formally express the dependencies among multivariate event data and reason over the different system states over time is of high importance in order not only to provide accurate predictions about future system behaviour but also to facilitate the supervision of the system by experts. Typical probabilistic logics [27] [28] provide the basic tools for the representation of the stochastic internal dynamics of event time series. Specifically, such tools allow reasoning over uncertain data by annotating ground facts with a probability value (i.e., probabilistic ground facts). Though, most of the times, this is not sufficient for expressing temporal correlations among the observed patterns. In order to represent uncertain data and time dependencies, probabilistic temporal logic (PTL) programming paradigms have been proposed [29] [30] [31] [32]. Such approaches extend the syntax and the semantics of probabilistic logic programs by terms of allowing for reasoning about point probabilities over time intervals through the use of probabilistic temporal rules.

Without loss of generality, a simplified notation for the representation of a *probabilistic temporal rule* can be represented as $A \rightarrow B : [t, p]$, where $A, B$ are (ground) formulae consisting of (ground) atoms and typical logic programming operators for conjunction, disjunction and negation ($\wedge, \vee, \neg$, respectively) while the (ground) formula $B$ is annotated with a probability value $p$ and a time unit $t$. Intuitively, such rule states that if the formula $A$ included in the *body* of the rule stands true at a fixed time, then the *head* formula $B$ will be also true with probability $p$ *after* $t$ time units. Similarly, we can use the same formalism to state that formula $B$ will be true *within* $t$ time units (i.e., use time intervals). Practically, the above type of logical expressions allows us to express logical operations over atoms that result in the probabilistic ground truth of other atoms within specific time.

Additionally, PTL programs allow for the use of *integrity constraints* that should not be violated across the execution time of the logic program. Specifically, two types of integrity constraints can be considered [29]: *block-size* constraints (BLK) are used to state that an atom $A$ cannot be consecutively true for more than a number of times that is defined as an integer value within the constraint expression; similarly, *occurrence* constraints (OCC) are proposed to state that an atom $A$ must stand true for a number of times that is within a range of minimum and maximum number of times value.

In the case of multivariate event data streams, we can imagine a system where a change in each event stream must not occur for more than $k$ times in a row. In this case, an expert responsible for the monitoring operation of the system could assert a set of BLK integrity constraints, one for each event stream separately (e.g., in case of $k = 3$, BLK($A$) :< 4, BLK($B$) :< 4, etc.). Additionally, we can consider a case of a system where a change in a specific event stream should occur at most one time. Similarly, the system expert could assert this information into the system in the form of an OCC integrity constraint (e.g., OCC($A$): $[0,1]$). Such kind of expressiveness provided by the support of integrity constraints could be useful not only for facilitating the supervision of a system by experts but also for refining the prediction of upcoming event sequences. For example, if there is a background knowledge that no more than 3 of the total $n$ events may occur at the same time step, a set of BLK constraints of the form BLK($ABCD$) :< 1, BLK($BCDE$) :< 1, etc. can be used to prune the number of considered states at each step.

In this example, each probabilistic temporal formula may represent a different state of the stochastic process and the prior probability of this state in the respective time step. We can also assume that the concurrent occurrence of all the three types of events may not be considered by adding a BLK constraint. Obviously, the rules contain only single atoms in their body since the considered model is a first-order Markov chain, i.e., subsequent states depend only on the current system state. In such scheme, additional rules for the computation of next step probabilities can be also added.

## 6. Event filtering

So far, we have discussed transparent ways to represent event dependencies independently of the underlying correlation scheme. An additional step in real- and near real-time event management considers the filtering of the derived dependencies. More precisely, correlation of events provides precious information for the behavior of the system over time. However, most sensor networks are highly dynamic by nature since they target to capture changes within non-static environments and spaces. Hence, the dependencies resulted in each step could be altered over time. In this context, a limitation of typical event correlation methods is their capability to identify *outdated dependencies*. For example, rules extracted from a correlation scheme based on Markov-models could reflect dependencies between events that took place a long time ago and do not conform with the recent activity of events. On the contrary, a memory-less scheme (e.g., one based on a sliding window) takes into account only the very recent activity by completely forgetting past event occurrences.

A simple, yet effective, approach for answering the question of when an extracted rule (i.e., dependency) becomes too old is to apply a time-dependent *aging* or *decay function*. In general, aging functions constitute time-based forgetting mechanisms. The main idea is that a drift on the behavior of a data stream should be a gradual process. Also, rules that have been extracted recently should be more important than the old ones. In the same context, the importance of a rule should decrease over time.

We represent an aging function by $\lambda(r_t) = f(t)$, where $r_t$ is a rule extracted $t$ time steps before. At each time step, each rule $r_t$ is assigned with an aging value (*weight*) depending on how old the rule is. The weight of each rule represents the importance of the rule at the current time step. In general, different $f(t)$ may be used depending on the particular situation for which events are being processed. Here, we discuss a linear and an exponential approach.

A linear aging function scheme with $\lambda(r_i) = -\frac{2k}{n-1}(i-1) + k + 1$ is presented in [33] where $n$ is the number of the considered past steps, $i$ is a counter starting from the current step and goes back over time, $r_i$ is a rule derived $i$ time steps before and $k \in [0,1]$ is the percentage of decreasing the weight of a rule by each step. By varying $k$ the slope of the aging function can be adjusted. Similarly, an exponential aging scheme with $\lambda(r_i) = \exp(-ki)$ is presented in [34]. The parameter $k$ controls how fast the weights decrease over time. For larger values of $k$, less weight is assigned to the rules. If $k = 0$ all the steps are assigned with the same weights.

Fig. 7 and 8 depict the above linear and exponential aging functions parameterized with distinct $k$ values.

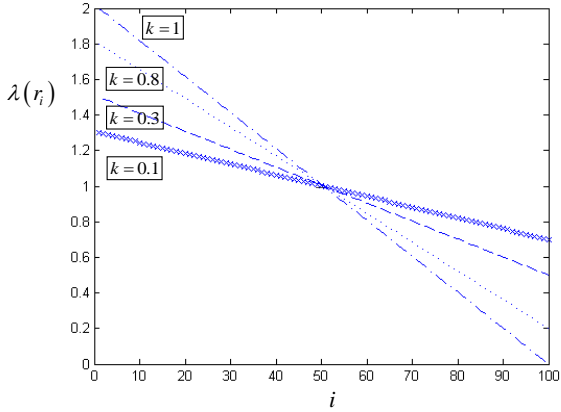
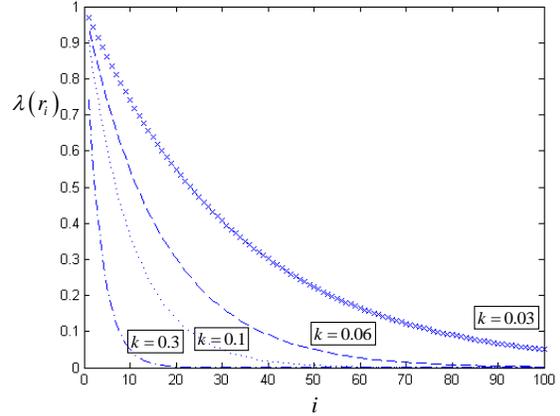

**Fig. 7.** Linear aging, $k = 0.3, 0.5, 0.8, 1$, $n = 100$.

**Fig. 8.** Exponential aging, $k = 0.03, 0.06, 0.1, 0.3$, $n = 100$.

If a rule that represents exactly the same dependency among events is generated in multiple time steps, the proposed framework balances between the, possibly varying, probabilities of rules by weighting the respective probability values according to the weights of the rules over time. Specifically, we monitor the rules extracted within the last $\Delta t_{mem}$ time steps and we use the following weighted form to estimate the probability of a rule derived within this time interval:

$$p_{r,\Delta t} = \frac{\sum_{i=t-\Delta t+1}^{t} \lambda(r_i) \cdot p_{r,i}}{\sum_{i=t-\Delta t+1}^{t} \lambda(r_i)} \quad (3)$$

Consider a case where $\Delta t_{mem} = 3$ and a rule $r$ that was extracted before 2 time steps with probability $p_{r,t-2} = 0.8$ and at the current time step with probability $p_{r,t} = 0.3$. Following Eq. (3) and taking into account the linear aging function with $k = 0.8$, the rule will be, finally, assigned with a probability $p_{r,\Delta t} = \dfrac{(0.8 \cdot 1) + (0.3 \cdot 1.8)}{2.8} = 0.47$. This way, a memory-less correlation scheme can be equipped with memory capabilities within a time interval $\Delta t_{mem}$.

## 7. Results

This section presents the results that have been derived during the evaluation phase of the proposed event management approach. All the experiments were performed by taking advantage of real data coming from the maritime domain. Specifically, our scheme has been assessed in a real-world scenario where the machinery and the infrastructure of three (moving) ships were constantly monitored making this information available for further analysis and processing at the back-end in real-time. The monitored sensor streams provided more than 3.5MB of data per minute resulting in a total of more than 450GB in less than 3 months (for all the three ships). CUSUM and Shewhart algorithms were used to detect possible abnormalities in the sensor streams. For each numeric stream, the result of the change detection process was a binary stream that was provided as input to the event correlation algorithm. We performed several tests to measure the precision of the algorithms. Each precision value was measured in 8000 sequential time steps. At each step, the predictions for the current time step have been checked of whether they became valid and the precision value was updated accordingly.

Fig. 9, 10 and 11 demonstrate the performance of the prediction scheme over different probability thresholds of the derived rules over time. The probability threshold value is an indication of the strength of each rule. Specifically, values close to '1' indicate that the rule was, in general, valid in the recent past while values close to '0' indicate obsolete rules. Hence, whenever a rule derived during the event prediction and filtering phases comes with a probability lower than this threshold, this rule has been discarded due to its low strength. Otherwise, it is inserted in the pool of rules that are evaluated based on their accuracy of predictions.

A first and clear result that can be deduced from the figures is that Shewhart detection dominates over CUSUM algorithm in all cases. The main reason behind this result is that CUSUM assumes that the underlying probability distribution of the dataset is a normal one. This is not the case for Shewhart which makes no assumptions with respect to the dataset. As a result, CUSUM fails to provide accurate detections that can lead to increased levels of precision with respect to future event forecasting. Apart from that, Fig. 9 depicts that CUSUM also fails to comply with high probability thresholds since the overall precision of results decreases as the probability threshold increases. On the other hand, as the probability threshold increases, the system improves its performance in terms of predictions in the case of Shewhart change detection, as it was expected.

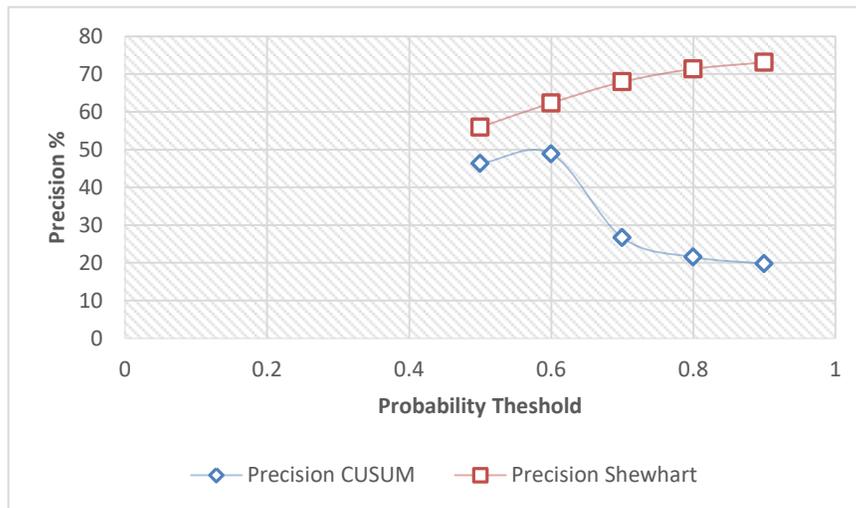

**Fig. 9.** Predictions with varying *probability threshold*, *m=l*=1.

Similarly to Fig. 9, Fig. 10 demonstrates the capability of the system to predict events that will occur two steps beyond based on events occurred during the last time step only. Compared to the case where the predictions referred only to the next step (Fig. 9), the quality of results is obviously lower (i.e., lower levels of precision). For instance, in the Shewhart case of Fig. 10, the precision value ranges between 47.8% and 57.0% while the respective curve in Fig. 9 ranges between 56.2% and 71.8%. Again, results are better in the case of Shewhart detection. This becomes more clear in Fig. 11 ($m=2$, $l=1$) where the precision results based on CUSUM detection do not exceed 10% in any of the considered cases.

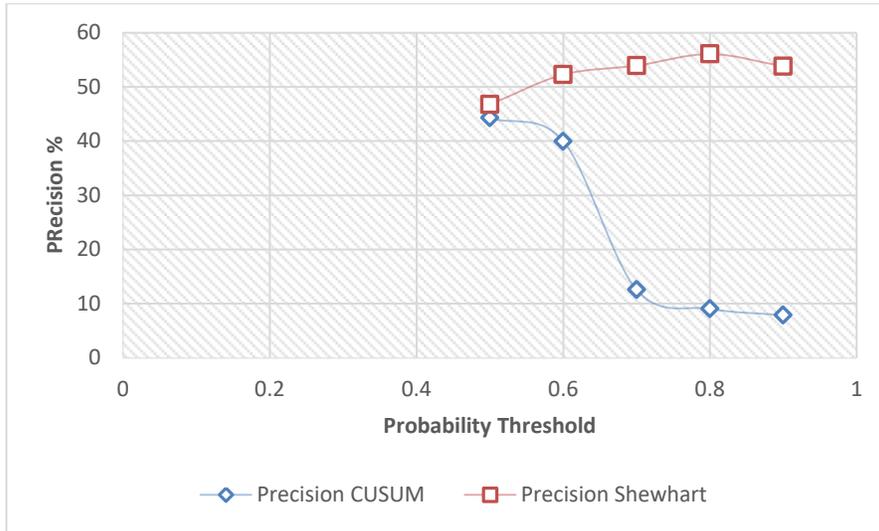

**Fig. 10.** Predictions with varying *probability threshold*, $m=1$, $l=2$.

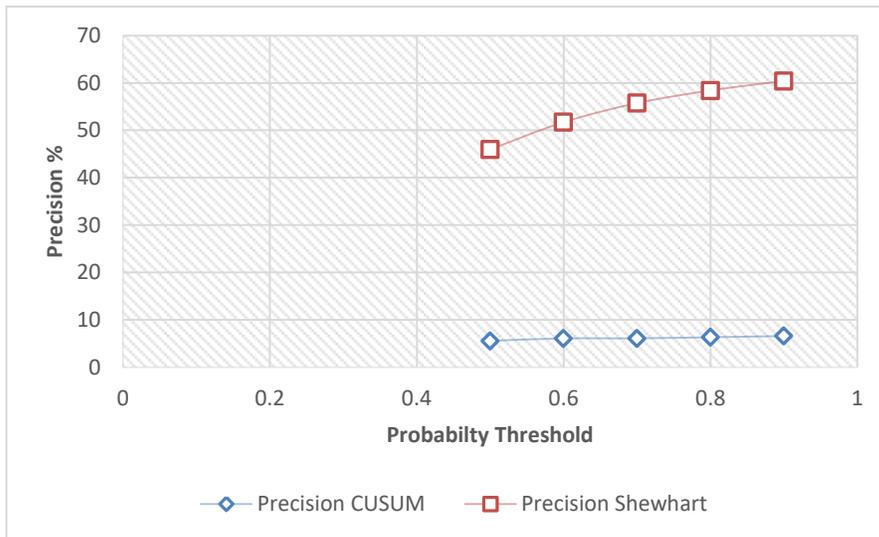

**Fig. 11.** Predictions with varying *probability threshold*, $m=2$, $l=1$.

We also assessed the quality of future event predictions based on different values of the fixed combinations parameter (i.e., $k$). Fig. 12 demonstrates the precision of the event predictions under different values of $k$, $l$, and $m$ when Shewhart algorithm was used to detect changes in the streams. No aging function was adopted in the experiment and the probability threshold was fixed to 0.6. It becomes clear that the accuracy of results is higher in the case where the event correlation algorithm is based only on the events that took place in the last step and not on the previous ones. Specifically, the 65.8% of predictions of the 'next step' events were valid when $k=1$ while the results were almost the same for $k=2,3,4,5$. It was also true that a higher number of steps considered to support a prediction (i.e., higher $m$ values) do not necessarily result in more accurate predictions. Moreover, in all cases, we measured lower precision values as the $k$ value was increasing. This practically means that most of the events can be deduced by taking into consideration only a single event that took place in the previous time step.

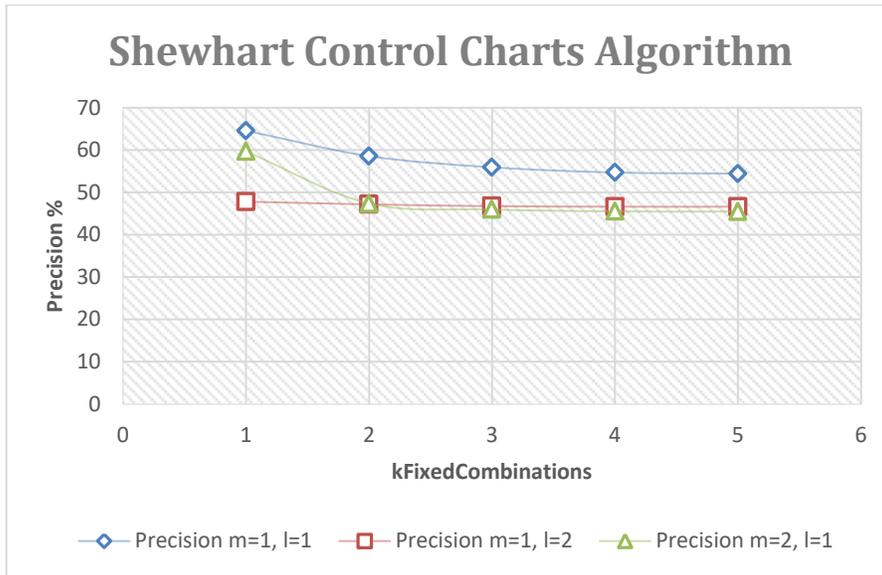

**Fig. 12.** Predictions with varying $k$ based on Shewhart detection.

Similarly, we assessed the performance of the predictions using the CUSUM algorithm for feeding the event correlation scheme (Fig. 13). As expected, the results were worse in the general case compared to the ones acquired using the Shewhart algorithm. This can be easily explained by the fact that CUSUM assumes a normal probability distribution for the underlying numeric stream and, again, none of the datasets comply with this requirement. As a result, the precision values were much lower than in the Shewhart case. For instance, all the precision results in the case when the two last steps were considered in order to predict a single step fall down to less than 25% of accuracy. The curve that was affected less by the underlying change detection algorithm was the one with $m=l=1$. Again, no aging policy was adopted and the probability threshold was set to 0.6.

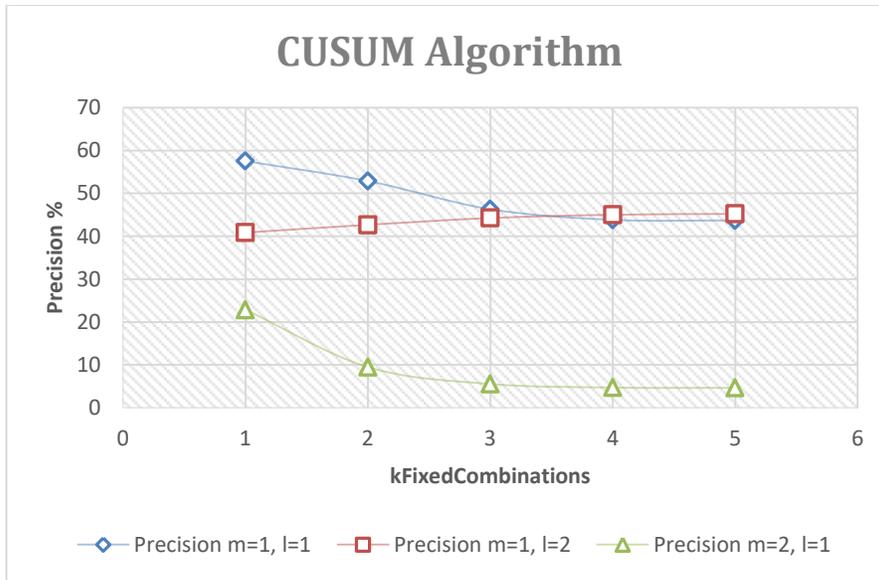

**Fig. 13.** Predictions with varying *k* based on CUSUM detection.

Additionally, we assessed the performance of the aging functions for both detection schemes considered over time. Fig. 14 and 15 present the system performance in the case where linear and exponential aging, respectively, was adopted. In both cases, it seems that the overall accuracy of predictions is enhanced compared to the case where no aging function is applied. Once more, Shewhart detection performed better than CUSUM in both cases because of the abnormality in the probability distribution of the data. Specifically, the overall precision on the prediction in the case of Shewhart detection was always more than 75% while it did not exceed the value of 45% in the CUSUM case. Moreover, Shewhart detection performed better in linear aging than it did in exponential aging. On the other hand, CUSUM performed better in exponential aging than in linear aging. In all cases the *k* value does not seem to significantly affect the overall quality of results. The probability threshold was set to 0.7 while *m=l*=1.

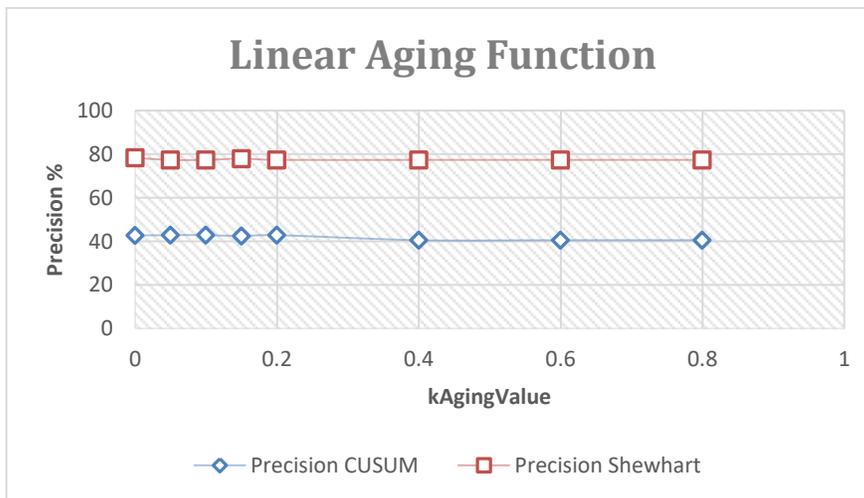

**Fig. 14.** Predictions with varying *k* based on linear aging.

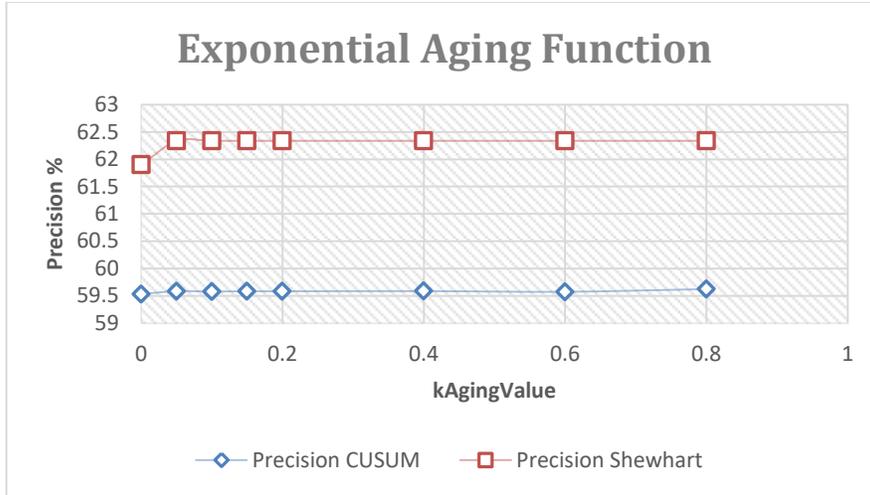

**Fig. 15.** Predictions with varying *k* based on exponential aging.

## 8. Conclusion

The paper discussed several parts of the processing chain in the context of online event management over sensor data. First, we demonstrated that change detection techniques can be used for the determination of events in sensor networks. Then, we proposed an online scheme for correlating multivariate event data. The algorithm involved a variable-order Markov model to capture sequences of multiple events. We also showed that representation of dependencies in order to predict future event sequences can be accommodated within a probabilistic temporal knowledge representation framework that allows the formulation of probabilistic rules. To address the important issue of outdated dependencies, we have set-up a time-dependent framework for filtering the extracted rules over time through aging factors. The experimental evaluation of the presented approaches was based on real-world data from the maritime domain.

The presented correlation scheme faced the problem of correlating multivariate event data by, essentially, adopting high-order Markov models to represent frequent event patterns. Since these models capture sequences of events that take place in subsequent steps, they have certain limitations in representing direct dependencies among events that occur within a larger timeframe. In most cases correlated events may not happen one after the other since the effects of the causal may need time to be observed. For example, we can consider the case where an event of type A which is coupled with a temperature sensor occurs exactly ten time steps after the occurrence of event type B related to a smoke detector. In that case, even if a ten-order model could capture this dependency through multiple time steps, it could not provide a direct correlation between the cause (i.e., increase of temperature) and the effect (i.e., existence of smoke). As a next step, we plan to investigate approaches that can address time dependencies among events within larger timeframes. A possible alternative could take advantage of a sliding window that would add memory to the algorithm.

Evidence of a bad or malfunctioning behavior of a sensor network is often embedded within sequences of events detected in distributed nodes across the network. Such state of importance may involve multiple time steps in order to reveal its effects. Also, several bad states could lead to the similar types of events. Our future research agenda will focus on the identification of hidden (i.e., unobservable) system states that may affect the observable variables (i.e., sensors). Hidden Markov Models set up a modeling framework for the representation of such causalities and the determination of unobservable states of a system with regard to the observed measurements.

Another aspect of importance is the improved characterization of detected changes in the context of a time series. Most of the existing change detection algorithms do not provide further annotation of the detected novelties (i.e., 0/1 characterization). We plan to extend existing univariate change detection techniques in order to provide multi-level detection of changes for numerical attributes. This step involves the identification of novel value ranges (i.e., clusters) with new entry values that do not comply with the existing clusters.